\documentclass[preprint2]{aastex}
\usepackage{psfig}

\def\beq{\begin{equation}}
\def\eeq{\end{equation}}
\def\bey{\begin{eqnarray}}
\def\eey{\end{eqnarray}}
\def\kms{\mbox{\rm \,km\,s}^{-1}}
\def\snia{SN Ia~}

\def\pc{\;\rm pc}

\def\msun{\;{\rm M}_\odot}

\def\spose#1{\hbox to 0pt{#1\hss}}
\def\lta{\mathrel{\spose{\lower 3pt\hbox{$\sim$}}
    \raise 2.0pt\hbox{$<$}}}
\def\gta{\mathrel{\spose{\lower 3pt\hbox{$\sim$}}
    \raise 2.0pt\hbox{$>$}}}

\input epsf

\begin{document}

\title{Dynamical limits on galactic winds, halo machos and intergalactic globular clusters}
%
%
\author{HongSheng Zhao
\footnote{Visiting professor at
Institute of Theoretical Astrophysics, National Astronomical Observatory, 
Beijing 100012, China (Email: hsz@ast.cam.ac.uk)}}
\affil{Institute of Astronomy, Cambridge, CB3 0HA, UK}
%
\label{firstpage}

\begin{abstract}
We argue that any violent galactic winds following early epoch of star
bursts would significantly weaken the potentials of galaxies, and
leave lasting signatures such as a lowered dark halo density and
preferentially radial/escaping orbits for halo tracers such as
globular clusters.  A galaxy is disintegrated if more than half of its
dynamical mass is blown off.  The presence of dense halos in galaxies
and the absence of intergalactic/escaping globulars should imply an
upper limit on the amount of baryons lost in galactic winds around
4\% of the total mass of the galaxy.  This
translates to limits on the baryons participating the early star
bursts and baryons locked in stellar remnents, such as white dwarfs.
The amount of halo white dwarfs claimed in recent proper motion
searches and microlensing observations in the Galactic halo are too
high to be consistent with our dynamical upper limits.  
Similar arguments also imply upper limits for
the amount of neutron stars and stellar black holes, in galaxy halos.  
Nevertheless, a milder outflow is desirable,
especially in dwarf galaxies, both for lowering their cold dark matter
central density and for injecting metals to the intergalactic medium.
\end{abstract}


\onecolumn

\section{Introduction}

How strong is the energetic feedback from star formation in galaxies?
What are the effects of massloss from a galaxy?  
These are important questions in understanding galaxy formation.
Baryonic gas cools and forms stars in potentials
of the dark matter halos.  Some massive stars explode as supernovae
and the stellar envelope is ejected with high speed.  This feedback
not only injects and mixs metals in galaxies, but also drives galactic
winds to pollute intergalactic medium.  A sudden loss of baryonic gas
by winds after an extremely powerful burst of star formation also
means a sudden weakening of the gravitational potential of the galaxy.

The weakening of the potential has several interesting consequences.
For example, the galaxy halo will relax to a looser distribution after
a mild loss of its mass through winds.  This reduction of halo density
is interesting for overcoming the dense cusp in Cold Dark Matter halos 
(e.g. Gnedin \& Zhao 2001).  Stars could also escape a lowered potential
well of a galaxy, and appear as intergalactic stars.

There are also limits on the amount of massloss.  A galaxy becomes
unbound if more than half of its mass is lost (Hills 1980).
Real galaxies appear to be tightly bound.  The boundness of
present day galaxies should set some limits on the violentness
of massloss and star formation in the past.  

It is now well-established that galactic outflows are ubiquitous in
actively star forming galaxies, both in the local universe and at high
redshift (see a review by Heckman 2001).  The amount of gas lost
in the energetic outflow is largely comparable to the amount of stars
formed.  Such a wind can be efficient in polluting
the hot gas in the intergalactic medium (IGM).
Some amount of energetic wind or outflow seems mandatory to account for
the presence of metals in the QSO absorption systems at high
redshift, and metals in the hot gas in galaxy clusters 
(e.g., Bookbinder et al. 1980).

In the Milky Way there are very few metal-poor stars with $[Fe/H] \le
-4$ in the halo and $[Fe/H] \le -0.6$ in the disk.  There are also
metals in some distant high-velocity clouds in the Local Group.  These
again suggest an early phase of metal production in a burst of mainly
massive stars.

Several recent claims of detection of a substantial amount of white
dwarfs (WDs) in the Galactic halo and distant galaxies by direct
proper motion searches or indirect microlensing observations suggest a
first burst of high-mass stars narrowly peaked around $(2-3)\msun$ in
galaxy halos.  To leave behind every half-solar-mass
WD, a few times more of the remnant mass is returned to the
interstellar medium via the planetary phase.  Those in binaries die
after $\sim 1$ Gyr via the highly energetic Type Ia SN, which can
easily power a hot and fast wind and drive all stellar ejecta out of
the potential well of a galaxy.  

Motivated by these considerations, we model the dynamical effect of
the post-star-formation galactic wind on the density distribution of
the dark particle halo.  We examine the effect of a fast wind at a
redshift $z \sim 1$ when the Milky-Way-sized halos are already largely
assembled, and energetic \snia are observed.  In general, a galaxy
will expand, stellar orbits become highly radial in response to a
rapid weakening of the potential well due to the mass gone with the
wind.  In particular we set limits on the amount of high-mass star
formation in galaxies in the past, and the amount of stellar remnents,
e.g., white dwarfs, in present day galaxies.

Simple models for the wind and expansion are developed in \S2 and the
Appendix, including the effects on globular clusters.
Applications to the halo white dwarf problem are made in \S3.  We
discuss our results in \S4, in particular, the strength and starting
time of the wind, and conclude in \S5.

\section{Models}

\subsection{The post-wind halo expansion and escaping stars}

Let us define $\epsilon$ as the fraction of the initial galaxy total mass 
that is lost in the wind.  It is then given by
\beq\label{epsdef}
\epsilon \equiv {M_w \over M_w + M_0},
\eeq
where $M_0$ is the present day total mass bound to the galaxy,
and $M_w$ is the amount of mass lost in the wind.
Define the constant $\epsilon_\infty$ to be the critical amount of massloss 
for the system to expands to infinity or unbound, then 
for a wind of a range of speed we have
\beq\label{fastslow}
{\rm [fast~wind]}~0.5 \le \epsilon_\infty \le 1~{\rm [slow~wind]},
\eeq
where the boundaries are specified such that 
a system unbounds if lossing 50\% of its mass in a fast wind
or 100\% of its mass in a slow wind.

The loss of gas reduces the gravitational mass of the system, making
the potential shallower. The system will relax to a new equilibrium by
an overall expansion.  The orbits of tracer particles in the galaxy
potential will also expand.  The expansion factors depend on the speed
of the massloss, and are derived in Appendix A.1 and A.2.  The weakening
of the potential well also means some stellar particles could escape.
The process is different from the massloss of the gas, which is pushed
out by the pressure gradient. Here stars at the high energy tails
could gain sufficient potential energy and exceed the bounding
potential of the system.  This fraction is estimated in Appendix A.3.

In general a system of initial characteristic size $r_{i,e}$ 
expands, after lossing $\epsilon$ fraction of the total mass, 
to a size $r_{f,e}$ with 
\beq\label{X}
X \equiv {r \over r_i} = 1 + {\epsilon_\infty \epsilon \over
\epsilon_\infty-\epsilon}.
\eeq
Note $X \rightarrow \infty$ when $\epsilon \rightarrow \epsilon_\infty$.
For a slow wind defined by $\epsilon_\infty=1$ we have $X=(1-\epsilon)^{-1}$.

The weakening of the potential due to massloss and overall expansion
has also the effect of increasing the apocenter of a tracer orbit.  
A tracer particle initially on circular orbit of size $r_i=a_i$ will 
expand to a generally eccentric orbit with an apocenter $r=a$, where
the expansion factor
\beq\label{A}
A \equiv {a \over a_i} \approx {1 \over 1-{\epsilon \over \epsilon_\infty}},
\eeq
so in a fast wind \footnote{For a slow wind we have $\epsilon_\infty=1$, hence
$A=X=1/(1-\epsilon)$ (cf. eqns.~\ref{X} and~\ref{A}), consistent with the 
expectation that the eccentricity of the orbit is constant for 
adiabatic changes of the galaxy potential (Lynden-Bell 1963)}
the expansion in apocenter is somewhat stronger than for the system as a whole,
i.e., $A ={a \over a_i} \ge {r \over r_i}=X$.
This is because 
the pericenters are determined by the angular momentum barrier, which
does not change very much, so the apocenter must increases faster than
the semi-axis.  As a result, orbits become preferentially radial.  
Note also the size of a tracer orbit doubles, i.e., $A=2$ for
$\epsilon=\epsilon_{\rm dbl}=0.5\epsilon_\infty$.

Another effect of weakening of the potential is that some stars or globular
clusters can escape the potential.
Let $\epsilon_*$ be 
the fraction of escaping stars among all stars produced in the galaxy, then
(cf. Appendix A.3)
\beq\label{esc}
\epsilon_* \equiv {M_{*, \rm esc} \over M_{*, \rm esc} + M_*} \approx 
\left[{(1-\epsilon_\infty)\epsilon \over (1-\epsilon)\epsilon_\infty}\right]^2.
\eeq
Note that no stars escape in a slow wind with $\epsilon_\infty \rightarrow 1$,
and all stars escape in a wind of maximum strength 
$\epsilon \rightarrow \epsilon_\infty$.  If the massloss is small, then
$\epsilon_* \sim \epsilon^2$ in a fast wind with $\epsilon_\infty=0.5$.

\subsection{Absence of intergalactic globulars and the maximum massloss}

We are primarily interested in fast winds with $\epsilon_\infty=0.5$.  
We shall also assume that the expansion factors $X$ and $A$ and
the massloss fractions $\epsilon$ and $\epsilon_*$ are all everywhere the same.
These are illustrated in Fig.1
as functions of gas massloss fraction $\epsilon$.
Since galactic winds are generally expected to be fast,
the above calculations motivate us to rate the degree of expansion caused by 
a fast wind into three categorical classes,
\begin{eqnarray}\label{mild}
\epsilon  & \sim & 0.04~~{\rm mild},\\\label{double}
          & \sim & 0.15~~{\rm severe},\\\label{unbound}
          & \sim & 0.5~~{\rm unbound}.
\end{eqnarray}
To justify these categories, 
note when a galaxy losses $\epsilon=15\%-25\%$ of its mass in a fast wind,
the galaxy expands in volume by a factor $X^3=1.8-3.4$, and the apocenters of 
the globular cluster tracers could increase by a factor of $A \sim 1.4-2$,
and a fraction $\epsilon_*=3\%-10\%$ of the galactic stars or globular
clusters should escape.  For a bright galaxy like the Milky Way or
the Andromeda with of order 150-450 globulars, 
this would predict of order 5-45 ``intergalactic'' globulars. 

Fig.2 shows the most recent compilation of
orbital parameters of the Milky Way globular clusters (Dinescu et
al. 1999 and references therein).  The distribution appears similar to
that of an isotropic models, which predict a median apo-to-peri ratio about
$4$ (van den Bosch et al. 1999).  Globular clusters with
apo-to-peri ratio $>10$ are rare, and these outliers appear as frequently
for inner disk globulars as for outer halo globulars;
the outer globulars should experience much less
dynamical friction and tidal destruction than the inner globulars.
There are very few Galactic globulars beyond 20 kpc of the Galactic center.
The total number of globular clusters in the Local Group is
estimated to be 
\beq
N_{\rm MW}+N_{\rm And}\sim 150+450=600
\eeq
(e.g., Barmby \& Huchra 2002),
and the numbers might have been much higher earlier on 
considering the destruction of inner globulars once the disk and bulge form.
The fact that there are no direct evidence for free floating clusters 
in the Local Group suggests {\it at face value} 
\beq
\epsilon^2 \sim \epsilon_* \le {1 \over 600},
\eeq
And as far as we know there are no direct evidence
for intergalactic globulars around other galaxies, implying perhaps
that they are very few.
Together these imply that the massloss in galaxies
in the past is limited to a fraction 
\beq
\epsilon \le 0.04.
\eeq  
We conclude the lack of escaping old halo globular clusters
in galaxies does not support a rapid expansion of galaxy halos.  

\section{Application to the halo white dwarfs}

\subsection{The observational and theoretical case for halo white dwarfs}

Recent searches for baryonic dark matter found between (2-5)\% to
(8-50)\% of the dark halo density to be in the form of
half-solar-mass WDs.  The lower estimates come from direct detection
of faint blue high proper motion objects in the solar neighbourhood
(Oppenheimer et al. 2001, Ibata et al. 2000), and the upper estimates
are indirectly inferred from observed microlenses towards the LMC
(Alcock et al. 1997, 2000, Lasserre et al. 2000).  While these observations
confirms the notion that the bulk of the halo is still in the form of 
undetected non-baryonic weakly interacting particles (WIMPs), 
the existence of a significant halo WD population is still very contentious
(see a recent review by Richer 2001).  
In fact the infered amount of WDs is surprisingly big, and is difficult 
to fit in current theories of star formation and galaxy evolution.
For example, the mean baryons to WIMPs split in the universe
is only around $15:85$ for current favored cosmological parameters.
Nevertheless, theorists have long been prepared to include this odd
piece in the cosmic puzzle (Carr 1994).  The most natural way to
account for these WDs and their extragalactic counterparts, albeit not
without difficulties, is that they are the relics of the first burst
of high-mass stars in all galaxies.  

The detailed properties of the
burst of WD progenitors must conform with a number of observational
and theoretical constraints.  The progenitors are likely formed about 10 
Gyrs ago, earlier than the disk, so that the hydrogen-rich WDs have
enough time to cool to a temperature of 3000-8000 K after their
progenitors evolve off the main sequence in 1-2 Gyrs (Hansen 1999);
the helium-rich WDs can cool faster, and are perhaps as
numerous as the hydrogen-rich WDs.  The progenitors must have
a narrow top-heavy IMF within the range of $1-8\msun$, perhaps a
log-normal distribution peaked around $2-3\msun$, so to avoid
overpredicting the counts of faint main sequence stars in the halo
(Chabrier, Segretain \& M\'era 1996 and references therein).  This
burst of WD progenitors must have a total mass at least comparable to
the total mass of galactic stars in the disk today, even using the
most conservative lower estimate of Oppenheimer et al. (2001).
Note we treat the formation of WD progenitor as a single burst, 
although some claim a spread in ages among the sample of white dwarfs 
from Oppenheimer et al. (Hansen 2001).

The bulk of the mass of a WD progenitor is in the envelope, 
which is injected back
into the interstellar medium at the end of main sequence lifetime
through the planetary nebulae phase.  The metals synthesized in these
progenitors are also released, which could severely distort the
abundance pattern of observed halo and disk stars, especially in
deuterium, helium, carbon and nitrogen, if the metals were kept inside
the galaxy.  This is saved by a galactic wind about 1 Gyr later
powered by Type Ia supernovae (\snia) evolved from binary progenitors
(Fields, Mathews, Schramm 1997, Fields, Freese, Graff 2000 and
references therein).

\subsection{A dynamical limit on white dwarfs}

Let us estimate the effect of a wind driven by such a burst of 
WD progenitors.  Let $M_0$ be 
the present day galaxy mass, $f_{WD}M_0$ be the mass locked in WD remnants.
For a progenitor mass of $m_{PG}$ to evolve
into a WD of mass $m_{WD}$, the amount of mass lost is $m_{PG}-m_{WD}$,
so the total number of WDs (as well as the WD progenitors) $N_{WD}$
and the mass lost in the wind is given by
\beq
M_w \equiv M_i-M_0 =  \left(m_{PG} - m_{WD}\right)N_{WD},
\eeq
and
\beq
N_{WD} = {f_{WD} M_0 \over m_{WD}},
\eeq
where $M_i \equiv M_0 + M_w$ is the pre-wind original total mass of 
the galaxy.
The fraction of the initial galaxy total mass that is lost in the wind,
$\epsilon$, is then given by
\beq\label{eps}
\epsilon \equiv {M_w \over M_w + M_0} = 
{\left({m_{PG} - m_{WD} \over m_{WD}}\right)f_{WD} \over
1 + \left({m_{PG} - m_{WD} \over m_{WD}}\right)f_{WD} }.
\eeq

Fig.3 shows the boundaries defined by eq.~(\ref{eps}) and 
eqs.~(\ref{mild},\ref{double},\ref{unbound})
in the parameter plane of $m_{PG}/m_{WD}$ vs. $f_{WD}$.  
For a log-normal distribution of $m_{PG}$ within $1\msun-8\msun$, 
we expect 
\beq
\left<m_{PG}\right>\sim 4 \left<m_{WD}\right> \sim (2.5-4.5)\msun,
\eeq
insensitive to the details of the IMF.
So, allowing only a mild wind and expansion, 
Fig.3 predicts that the maximum mix of halo WDs with halo WIMPs is given by
\beq\label{pred}
{f_{WD} \over 1-f_{WD}} \le {2\% \over 98\%}
\eeq
in all tightly bound galaxies.

\section{Discussions}

\subsection{Did the Galaxy experience a severe expansion?}  

Apart from the globulars, another constraint for the amount of 
evolution of the Galactic potential is 
that the present-day dynamical mass of the Milky Way out to a shell
at 50 kpc (the distance of the LMC) is about $0.5\times 10^{12}\msun$
(Wilkinson \& Evans 1999), implying an average density of 
$\sim 10^{-3}\msun{\rm pc}^{-3}$ inside this shell.
One can also estimate the density of an initial dark halo
of the Milky Way, averaged over the same radius.  
Assume the inner 50 kpc of the Galaxy is already virialized
at $z=2$, then it should have an overdensity of at least some 200 times
the mean density of the universe then, insensitive to cosmology.
More precisely a collapsed halo has a mean density
\beq
\rho_{\rm vir}(z) \sim {180 (2+\Omega_z) H_0^2 \over 8 \pi G} (1+z)^3 
\sim 0.8\times 10^{-3}\msun{\rm pc}^{-3},
\eeq
where $H_0=75$km/s/Mpc, $z=2$ and $\Omega_z \sim 0.9$ 
for $\Omega_0=0.3$ and $\Omega_\Lambda=0.7$ cosmology
(e.g., Bullock et al. 2001).
So the density of the dark halo at $z=2$ agrees with the present day 
halo within 50 kpc very well.  Another way to estimate this is with the  
Navarro, Frenk \& White (1996) or 
Moore et al. (1998) models, assuming a virial radius of the 
Milky Way halo (200 kpc-400 kpc), 
and the concentration parameter of 6-12.
These models can predict the mean density within $50$ kpc,
and this turns out to be $(0.5-2) \times 10^{-3}\msun{\rm pc}^{-3}$,
i.e., within a factor of two of current density.  
In comparison if a wind doubles the size the galaxy,
the mean density enclosed inside a shell drops by a factor of $X^3=8$.
This implies that any dynamical evolution must have been very mild
after the formation of the outer halo.

\subsection{When did the galactic wind start?}

The oldest halo globular clusters
are as old as about 12 Gyrs (Salaris et al. 1997), and the coolest
local disk WDs are $9-12$ Gyrs (Knox et al. 1999, Oswalt et al. 1996,
Hernanz et al. 1994).  It is plausible to assume that $9-12$ Gyrs ago
is roughly the epoch when baryons collapse in the potential of the
dark particle halo of the Milky Way, and start forming the first stars
everywhere, in the local disk, in globular clusters and in the halo;
the IMF is top-heavy in the outer halo and normal Salpeter elsewhere
in the disk and in globular clusters.  This epoch corresponds to a
redshift of $z\sim 1-3$ in a 14-Gyr-old universe with
$1-\Omega_m=\Omega_\lambda=0.66$ and $H_0=66$km/s/Mpc.  The Madau
diagram, i.e., the extragalactic star formation rate in this high
redshift window is still not yet well constrained.  
There is a delay between the onset of the halo star
formation and explosion of \snia, which is the time for the secondary
of a binary to evolve and overflow its Roche lobe, and bring the WD of
the primary into a critical mass.  Take a plausible mass of $\sim
(4-8)\msun$ for the primary, and $\sim (1-2)\msun$ for the secondary,
the delay is $\sim 1-8$ Gyr.  This predicts the appearance of the first
halo \snia $\sim 8-11$ Gyr ago,
which seems to be consistent with current detection of some \snia
at redshift $z=1-2$.  In short, we estimate that the
galactic wind powered by the first \snia in the halo starts at $z \sim
1$, or when the universe was about half its present age.  Since major
mergers are rare in Milky Way sized halos after $z = 2$ in 
the fairly successful $\Lambda$CDM cosmology, 
it appears safe to assume that CDM halos as big as the Milky Way halo are well
assembled through hierarchical merging and infall by the time of the
wind.

\subsection{Are the winds fast or slow?}  

A slow wind means nearly adiabatic expansion, which happens if the
outflow velocity is much less than the circular velocity of the
galaxy.  Observationally prominent outflows are seen in active star
burst galaxies such as M82 at low redshift.  Generally speaking the
outflow rates are comparable to the rates of star formation in terms
of mass and energy output, and these superwinds driven by star bursts
have velocities of $100-1000\kms$ (Heckman 2001).  A wind or jet must
also carry sufficient momentum in order to deliver and mix the metals
observed in the intergalactic medium.  At high redshift winds with
Doppler offset velocity of 200-1000$\kms$ have been seen in
Lyman-break galaxies with active star formation (e.g., Pettini et
al. 2002), suggesting for fast outflow driven perhaps by UV-photons
and SNe from the star burst.  It would be oversimplifying to associate
these observed star bursts directly with the speculated halo WD
progenitors, but it is perhaps safe to argue that a hypothetic burst
with a top-heavy IMF should lead to outflows at least as violent as
these observed ones.

We could also estimate the energy budget of winds theoretically, as in
Fields, Freese \& Graff (2000).  Ejecta from a \snia has a velocity
$v_{\rm ej} \ge \sqrt{10^{51}{\rm ergs}/8\msun} \sim 2500\kms$, 
where $10^{51}$ erg is the kinetic energy per \snia, and $8\msun$ is
the upper limit for the mass of the exploded star.  So the ejecta
could escape any giant galaxy instantaneously by its own kinetic
energy.  It would require fine tuning of the star formation efficiency
and/or the binary fraction to keep the outflow velocity low, at least
for a top-heavy IMF.\footnote{Using a bottom-heavy IMF can suppress
(binary) stars above $1\msun$ and the \snia, but 
it is not interesting for producing old halo WDs in the
first place.   The ejecta of low-mass stars drives a wind of velocity of 
$\sim 30\kms$, fast for dwarf galaxies.}  
As an illustration, consider a very inefficient
star burst early on with every $1000\msun$ of gas of a galaxy forming
$10$ single stars of $6\msun$ plus one binary of masses $4\msun:2\msun$.  
Assume this top-heavy IMF leaves behind a remnent of 
a $1\msun$ WD with another $5\msun$ of stellar
material ejected back to the interstellar medium with a velocity of
$30\kms$ for a single progenitor, and $3000\kms$ for a binary
progenitor.  Most part of the ejecta might escape the galaxy with a
high speed through ``holes'', however, if only 25\% of the ejecta could 
mix and thermalize with another $934\msun$ of unprocessed gas, then the mean
outflow velocity is still above $3000\kms \sqrt{25\% \times
5\msun/989\msun} \sim 100\kms$, i.e., the outflow velocity is still far
from adiabatic with a speed comparable to the velocity dispersion
(sound speed) for the Milky Way-sized galaxy, and exceeding the escape
velocity of dwarf galaxies.  So an adiabatic expansion in the Milky
Way would require ``tuning'' the binary formation efficiency $\ll 10\%$,
the star formation efficiency $\ll 6.6\%$ and 
the energy deposit efficiency $\ll 25\%$.  
Without this fine-tuning, we expect theoretically 
that the energy from the SNe is too high for a slow galactic wind.
\footnote{
It is a common misconception that pressure-driven winds have necessarily 
a sound velocity very close to the typical velocity dispersion
of the dynamical system.  Calculations of winds in globular clusters show
that the sound velocity can exceed 
the typical velocity dispersion and escape velocity of 
the dynamical system by a factor 1-10 if the energy injection rate is high
(see, e.g., Fig. 2 and Fig. 3a of Faulkner \& Freeman 1977).
The gas flow velocity can also exceed the sound velocity in some regions.}

In short, we conclude from theoretical and observational evidences
that the galactic winds have speeds  
somewhere between our slow wind and our fast wind limits, with a
tendency to be fast winds.  For simplicity we shall adopt the fast
wind limit in this paper, which could somewhat overestimate the
effects of the wind.

\subsection{Comparison with independent observations}

The most direct estimates for the WD fraction $f_{WD}$ come from
direct searches by Ibata et al. (2000), and by Oppenheimer et
al. (2001); the latter claims a firm lower limit from their large
sample of high proper motion halo WDs, about $38$ with spectroscopy
confirmation and some show hydrogen lines, indicative of a mix of hot
and cold WDs with temperature in the range $3000-8000$ K.  Indirect
measurements come from accounting for $10-20$ microlensing events
towards the LMC from the MACHO group (Alcock et al. 1997, 2000) and
from the EROS group (Lasserre et al. 2000); the latter claims only an
upper limit on the basis of a null detection of a conclusive halo
microlens.  Clearly our theoretical model is consistent with the EROS
upper limit, but inconsistent with the best values from the MACHO
group in 1997 and 2000.  It is possible that most of the microlensing
towards the LMC are dominated by stellar microlenses in the LMC 
(Sahu 1994, Zhao \& Evans 2000) 
with only a minor contribution from the halo WDs.  Our
theoretical result, eq.~(\ref{pred}), which is effectively an upper
limit on any WDs in the halo, is consistent with the most
recent lower limit by Oppenheimer et al. (2001).

Our dynamical constraint is the tightest for the high mass progenitors.
One would hope that the progenitor mass could be constrained by the
temperature of observed WDs together with a cooling model and an
estimation of epoch of formation.  For example, 
the cosmic clock in Fig.3 indicates a 10-Gyr-old burst.
Their WD remnants 
are now hotter than $4000$ K if $m_{PG} \ge 2.2\msun$, and cooler
if $m_{PG}<2.2\msun$.  Unfortunately progenitors of $2\msun$ and $4\msun$
formed $9-12$ Gyrs ago both evolve to WDs in $\le 1$ Gyr, which then
cool to nearly the same temperatures after another $\sim 10$ Gyrs
(cf. Fig.4), so we cannot differentiate them.  We could, however, separate
the WDs of $1\msun$ progenitors from WDs of $4-8\msun$ progenitors
since the former will have a temperature (4000-10000 K) much higher
than the latter ($<4000$ K).
The temperatures of the observed halo blue WDs are spread out in the range
3000 - 8000 K (Oppenheimer et al. 2001).  This means there must have
been some $1\msun$ progenitors and some $8\msun$ progenitors,
consistent with a log-normal IMF peaked around $2-3\msun$ 
(Chabrier et al. 1996).

Fig.5 shows the dynamical limits on cool and hot WDs from
a 10-Gyr-old population.
Note stronger limits on cooler WDs because these have
more massive progenitors with higher $m_{PG}/m_{WD}$ ratio.
For a typical temperature of 4000 K, we find $f_{WD} \le 2\%$,
consistent with our earlier dynamical limit (eq.~\ref{pred}).

\subsection{The minimal dynamical feedback}

The dynamical feedback is generally strong.  
One way to minimize it is to form the WD progenitors in a thick disk instead
of in the halo.  It has been argued that some of the high proper
motion WDs in the Oppenheimer et al. (2001) sample have thick disk
kinematics (Reid, Sahu \& Hawley 2001, Koopmans \& Blanford 2001).  
Assuming all WDs and their
progenitors are in a thick disk of a scale height $1$ kpc, we can
convert the WD number density of $2\times 10^{-4}{\rm pc}^{-3}$ of
Oppenheimer et al. (2001) into a local column density of progenitors
of $0.4m_{PG}\msun{\rm pc}^{-2}$.  In comparison, 
the local column density of the Galactic thin disk
is $\sim (30-40) \msun{\rm pc}^{-2}$ within $|Z|< 1$ kpc, 
and the dark halo column density is $\sim (100-200)\msun{\rm pc}^{-2}$ 
within a vertical column $|Z|<\infty$ 
depending on the density profile of the halo.
We estimate $\sim (0.6-1.2)\%$ of the dynamical mass in the local column
is in WD progenitors if $m_{PG}\sim 3\msun$.  Take $\epsilon=2\%$ as
the characteristic fraction of the dynamical mass for the inner halo
lost in a wind, then the inner halo would expand by a factor $X \sim
1.02$ for an instantaneous wind.  It is interesting to compare this
with the dynamical effect of the formation of a galactic disk
collapsing out of the spherical halo.  The thin disk makes up $\sim
15\%$ of the dynamical mass in the local vertical column with
$|Z|<\infty$.  If a galactic thin disk with 15\% of the mass of the
halo form adiabaticly, the halo contracts by a factor about $0.85$.
So the feedback from a thick disk WD population is dynamically
unimportant compared to the formation of the galactic thin disk. 

\subsection{Formation of baryonic disks and mild wind}

How does the amount of mass lost in a mild wind compare with 
the total amount of baryons in a galaxy?  Take
a mean density of baryons in the universe
$\Omega_b=0.05(H_0/66)^{-2}$, and the fairly successful $\Lambda$CDM 
cosmological models with
$1-\Omega_m=\Omega_\lambda=0.66$ and $H_0=66$km/s/Mpc, we expect a
typical baryonic fraction in galaxies 
\beq
{\Omega_b \over \Omega_m} \sim 15\%.
\eeq
So we can express the wind massloss in terms of the total baryons
in a galaxy.  A mild wind (cf. eq.~\ref{mild}) looses 25\% of the total
baryons of a galaxy 
\beq\label{mild1}
M_w \sim 25\% M_b,\qquad M_b \equiv {\Omega_b M_i\over \Omega_m}.
\eeq

A mild wind is also relevant for star formations in galaxy disks with
a normal Salpeter type initial mass function with very few massive
stars.  Here the baryons of a galaxy cool and collapse adiabatically
to a disk.  When star formation ends, the envelope of evolved stars
are ejected out with a mild galaxy wind; the amount of stellar ejecta
$\sim 25\%$ of the initial stellar mass for a bottom-heavy IMF, so the
wind is mild.  We will not digress into this, except remarking two
counter-acting effects of disk formation on the density profile of
galaxy halo: the collapse of gas into disk stars leads to contraction
of the halo, and the later massloss from winds leads to expansion of
the halo.

\subsection{Disintegration of galaxies by winds and intergalactic stars and globulars}

What are the chances and consequences for a heavy wind to disintegrate
some galaxies in the past?  Systematic destruction of dwarf galaxies
would be desirable because of the notorious overprediction of dense
dwarf satellites in generic CDM models.  Another consequence of
disruptions is that the stars and their remnants, and perhaps globular
clusters can disperse with the unbound dark halo into the
intergalactic space and are later recycled to into the dark halo of
the second generation massive galaxies, such as the Milky Way.  Let
$f_b$ be the baryonic fraction in a galaxy, we expect in $\Lambda$CDM
models a typical baryonic fraction in galaxies $\left<f_b\right> \sim
\Omega_b/\Omega_m=15\%$.  To disrupt a galaxy, we need to have about
half or more of a galaxy to be in baryons, i.e., $f_b > \epsilon \ge
0.5$, where $\epsilon$ is the wind-lost baryons as a fraction of the
mass of a galaxy.  So a wind-disrupted galaxy needs to be very
baryon-rich, three times that of the universal value
$\left<f_b\right>$.  So the chances for disruption are likely low,
unless the universal baryonic fraction is much higher than $15\%$.

There are no strong evidences for intergalactic stars or globular
clusters although some have been claimed in galaxy clusters.  Both M87
and NGC1399 have excess star lights and numerous globulars around
them, and the outer globulars have velocities which are more
characteristic being bound by the galaxy cluster rather than the
central galaxy (e.g., Kissler-Pratig et al. 1999).  Theuns \& Warren
(1997) also discover 10 PNe in the Fornax cluster but unassociated
with any bright galaxies.  One complication of interpreting galaxy
cluster data is the projection effect: the PNe or globulars might be
``markers'' of halos of low surface brightness galaxies in the galaxy
cluster in the line of sight.  Even the bona fide free-floaters are
likely striped from spiral galaxies due to frequent merging in a
galaxy cluster (Moore et al. 1999), which is very different mechanism
from the wind in an isolated galaxy.  In any case the cD galaxies are
ambiguous systems for constraining galaxy winds compared to the Local
Group galaxies.

In principle some amount of free-floating globulars cannot be ruled out in the
Local Group.  When placed at 1 Mpc, the edge of the Local Group, 
a $10^5L_\odot$ globular would appear as bright as a K-dwarf star at 1 kpc
in the halo, but has a half-light radius of about $5\pc \sim 1\arcsec$.  
Much of the sky has not yet been covered by
detailed searches except for perhaps a few thousand square degrees:
along the Magellanic Stream ($10^o \times 100^o$) and the Sagittarius
Stream ($10^o \times 360^o$) and around Andromeda ($10^o\times 10^o$).
High quality data over 1500 square degrees has been released by 
the on-going SLOAN survey.  Assume 10\% of the Local Group is well surveyed
and no new globulars are found, then 
$\le 10$ globulars could be by chance at the
edge of the Local Group in unsurveyed area.  

\section{Conclusions}

In short, the absence of obvious signs of severe expansion of the
Milky Way and the absence of escaping globulars in galaxies in general
suggest that star formation and feedback in galaxies are mild;
galactic winds carry less than 4\% of the total gravitational mass, or
25\% of the baryonic mass in galaxies.  This restricts the amount of
remnents in galaxy halos from early star formation, and cool halo WDs
can make up 2\% of the total mass in present day galaxies.  This is
consistent with only the lower end of the halo WD fraction of $2\% \le
f_{WD} \le 50\%$ from direct and indirect detections of WDs; it is
consistent with Oppenheimer et al. (2001) detection, but inconsistent
with the higher values from Alcock et al. (1997, 2000) and Ibata et
al. (2000).  Our dynamical limit is also tighter for very cool WDs than for hot
WDs.  There is very limitted room to fit in a large extra population
of very cool WD population $\le 4000$ K, something that future deeper
surveys should take into account.  

Our model implies stringent limits on the amount of black hole (BH) 
or neutron star (NS) formation as well.  
Assuming these are remnents of massive stars formed in galaxy halos, 
then typically neutron stars or stellar black holes are 
formed from very massive ($8\msun-1000\msun$) progenitors with
$m_{PG}/m_{BH} \ge m_{PG}/m_{NS} \sim 6-30$.  So the massloss
is even more severe than for forming WDs.  The progenitors
evolve very quickly off the main sequence, and some explode
as Type II SN, which immediately power a very fast wind.
To keep the wind from severly damage a galaxy 
(cf. eq.~\ref{eps} and eq.~\ref{mild}) would imply 
that these remnents make up a fraction 
\beq
f_{BH} \le f_{NS} \le (0.1-1)\%
\eeq
of the total mass of a galaxy.
In comparison, the upper limit on stellar BHs from microlensing is 
$f_{BH} \le (30-100)\%$ (Lasserre et al. 2000, Alcock et al. 2001).
So our dynamical limit is much tighter.  
Some models of galaxy formations suggest a significant amount of massive black 
holes in galaxy halos, with black hole mass $m_{BH} \sim 10^2-10^6\msun$
(Lacey \& Ostriker 1985, Madau \& Rees 2001).  These black holes, however, must
be pregalactic, in which case our dynamical limits do not apply.

In summary, we show that intergalactic stars and globular clusters and
the present day density profiles of galaxies offer a new diagnosis of
early star bursts of massive stars in halos of galaxies, and the
amount of remnents, such as halo WDs and BHs.
A mild wind-induced expansion of the dark halos might
have played some role in lowering the dense cold dark matter
central density in dwarf galaxies (Gnedin \& Zhao 2001)
and high surface brightness galaxies as well.


HSZ thanks Bernard Carr, Rodrigo Ibata, Mike Irwin, Gerry Gilmore,
Donald Lynden-Bell, Jerry Ostriker, Tom Theuns and the anonymous
referee for helpful comments.

\appendix

\section{Appendix}

\subsection{Effects of massloss on the size of the galaxy}

A rapid change of the potential of a galaxy after massloss can cause
the galaxy to expand.  The final size can be estimated
with the virial theorem.  Let $M_i$ be the initial total mass, 
$M_f=(1-\epsilon)M_i$ be the final mass of the galaxy, and 
$r_{i,e}$ and $r_{f,e}$ be the initial and final effective sizes of the galaxy
such that
\beq\label{Ws}
W_i \equiv -{GM_i^2 \over 2r_i^e}, \qquad 
W_a \equiv -{GM_f^2 \over 2r_i^e}, \qquad
W_f \equiv -{GM_f^2 \over 2r_f^e},
\eeq
where 
$W_i$, $W_a$, and $W_f$ be the total potential energies of the galaxy 
before, right after, and long after the expansion.
Then the total energies of the system before and after the expansion are 
given according to the virial theorem by
\beq
W_i + {1 \over 2} M_i V_{\rm rms}^2 = {1 \over 2}W_i, \qquad 
W_a + {1 \over 2} M_f V_{\rm rms}^2 = {1 \over 2}W_f,
\eeq
where $V_{\rm rms}^2$ is the root mean squared
velocity of the system before or right after the instantaneous expansion.
Combining the equations we find the overall expansion factor (cf. Hills 1980)
\beq
\left.{r_f^e \over r_i^e}\right|_{\rm fast} = {1-\epsilon \over 1-2\epsilon}.
\eeq

In comparision, if the massloss is adiabatic, then
a shell initially at radius $r_i$ maps to another shell at radius $r$ 
adiabaticly conserving the angular momentum 
$\sqrt{G(M_0+M_w)r_i}=\sqrt{GM_0r}$, so the expansion factor
\beq
\left.{r_f^e \over r_i^e}\right|_{\rm slow} = {1 \over 1-\epsilon}.
\eeq

An interpolation of the equations for the fast and slow massloss
can be made to model massloss of arbitrary speed.
Generally the expansion factor $X$ can be modeled as a function
of the massloss fraction $\epsilon$ as following
\beq
X \equiv {r_f^e \over r_i^e} = 1 + {\epsilon \epsilon_\infty \over
\epsilon_\infty-\epsilon},
\eeq
where the constant $\epsilon_\infty$ is the critical amount of massloss 
for the system to expands to infinity $X \rightarrow \infty$.
For a wind of a range of speed we have
\beq
{\rm [fast~wind]}~0.5 \le \epsilon_\infty \le 1~{\rm [slow~wind]}.
\eeq

\subsection{Effects of massloss on the size of tracer orbits}

A rapid change of the potential of a galaxy after massloss can also alter
the orbital distributions of its tracers, e.g., the halo globular clusters.  
The effects on the apocenter are easiest to work out for the tracers on 
initially circular orbits.  A circular orbit of initial radius $a_i$ 
can become radial because of the reduced restoring force after the massloss
and reach a new apocenter $r=a$ while keeping the pericenter at $r=a_i$.
The new apocenter is determined by the specific energy equation,
\bey
E &=& {J_i^2 \over 2a_i^2} -\int_0^{a_i} {GM(r') \over r'^2} dr \\
  &=& {J_i^2 \over 2a^2}   -\int_0^{a}   {GM(r') \over r'^2} dr',
\eey
where the first terms are the azimuthal kinetic energy for an orbit
with the initial specific angular momentum $J_i=GM_i(a_i) a_i$,
the second terms on the r.h.s. of the equations
are final potential of the galaxy, determined by the 
final distribution of the mass 
\beq
M(r) \equiv (1-\epsilon)M_i(rX^{-1})
\eeq
in the galaxy, where 
$M_i(r_i)$ is the initial mass distribution inside a radius $r_i$,
and $X$ is the overall expansion factor.
Approximate the initial potential by a scale-free potential
with the mass profile given by
\beq
M_i(r_i) = \delta_i r_i^{\beta},\qquad 0 \ge \beta < 3,
\eeq
where $\beta$ is the powerlaw slope, and 
$\delta_i$ is a normalisation constant.  
Combining the above equations, we have
\beq
(1-\epsilon)X^{-\beta}  =  {(1-\beta) (1-A^{-2}) \over 2 (1-A^{\beta-1})}, 
\qquad A \equiv {a \over a_i},
\eeq
which relates between the expansion factor
of the apocenter $A=a/a_i$ and the massloss fraction $\epsilon$
in a fast wind; the factors $A$, $X$ and $\epsilon$ are the same everywhere
for the assumed scale-free potential.  For a Keplerian
potential with $\beta=0$, the above equation has a solution
\beq
\left.{a \over a_i}\right|_{\rm fast} = {1 \over 1-2\epsilon}.
\eeq
It turns out the above solution is accurate for 10\% for models with 
$\beta \ne 0$, and $0<\epsilon<0.25$.
This applies to non-scale-free potentials
such as the Navarro, Frenk \& White (1996) models as well
because the result is insensitive to the mass profile parameter $\beta$.  

In comparison, if the wind is slow, then the orbit will remain circular
and expand adiabaticly by the same factor $X$ as the galaxy, so
\beq
\left.{a \over a_i}\right|_{\rm slow} = X = {1 \over 1-\epsilon}.
\eeq

For wind of any speed, we can write 
\beq
A \equiv {a \over a_i} \approx {1 \over 1-{\epsilon \over \epsilon_\infty}}.
\eeq
Note $A \rightarrow X \rightarrow 2$ if 
lossing 50\% of the orginal mass in a slow wind ($\epsilon_\infty=1$), and
$A \rightarrow X \rightarrow \infty$ 
if lossing the same amount of mass in a fast wind ($\epsilon_\infty=0.5$).

\subsection{Effects of massloss on escaping stars}

A sudden weakening of the potential will boost the energy of some particles 
above their binding energy, and escape.  The average specific energy 
of the system before and long-after the massloss is related to the
characteristic potential energy (cf. eq.\ref{Ws}) by the virial theorem, so  
\beq
E_i = - {W_i \over 2 M_i}=-{GM_i \over 4r_{i,e}},
\qquad E_f = - {W_f \over 2 M_f}=-{GM_f \over 4r_{f,e}}.
\eeq
Substitute in the expansion factor, we have
\beq
{E_f \over E_i} = {M_f \over M_i} {1 \over X} = 1-2\epsilon.
\eeq
Particles in a King model or any smoothly truncated model should 
be bound between two energies, $\phi_{\rm min}$ at the bottom
of the potential well, and $\phi_{\rm max}$ at the outer truncation.
Long after the expansion, the truncation energy
$\phi_{\rm max}$ is increased to 
\beq
\phi'_{\rm max}=(1-2\epsilon) \phi_{\rm max}.
\eeq
During the sudden expansion, a particle at the bottom 
of the potential well will gain a potential energy
$-\epsilon\phi_{\rm min}$ without changing kinetic energy, hence
the particle's energy $E$ is increased to 
\beq
E'=E-\epsilon\phi_{\rm min}.
\eeq
So this particle could escape the final potential well if 
\beq
E' \ge \phi'_{\rm max}.
\eeq
This suggests that all particles with initial energy $E$ satisfying 
\beq
E-\phi_{\rm max} \ge  \epsilon (\phi_{\rm min}-2\phi_{\rm max})
\eeq
could escape the final potential, i.e., the escaping particles
occupy a narrow band at the upper end of the energy spectrum of the initial 
system, and the band width is $\propto \epsilon$.
The high energy tail of initial particles should follow
a distribution function $f(E)$ with
\beq
{dM(>E) \over dE} \propto f(E) \propto (\phi_{\rm max}-E)
\eeq
to have a smooth truncation.
So the amount of mass in the high energy tail $M(>E)$ is given approximately by
\beq
{M(>E) \over M(>E_{\rm min})} 
\approx {(\phi_{\rm max}-E)^2 \over (\phi_{\rm max}-\phi_{\rm min})^2}.
\eeq
Hence the escape fraction is given by
\beq
\epsilon_* \equiv {M_{*,\rm esc} \over M_{*, \rm esc} + M_*} 
\approx  \epsilon^2 
{(\phi_{\rm min}-2\phi_{\rm max})^2 \over (\phi_{\rm max}-\phi_{\rm min})^2}.
\eeq
For a highly concentrated system, e.g., a King model with a core radius
100 times smaller than the truncation radius, we have
\beq
|\phi_{\rm max}| \ll |\phi_{\rm min}|.
\eeq
So for highly concentrated systems we have
\beq
\epsilon_* \approx  \epsilon^2.
\eeq

The above estimations can be generalized to winds of any speed with the
following approximation
\beq
\epsilon_* \equiv {M_{*,\rm esc} \over M_{*, \rm esc} + M_*} \approx 
\left[{(1-\epsilon_\infty)\epsilon \over (1-\epsilon)\epsilon_\infty}\right]^2.
\eeq
This parametrization is justified because it reproduces nicely two simple 
limiting cases: 
(i) no stars escape in a slow wind with $\epsilon_\infty \rightarrow 1$;
(ii) all stars escape in a wind of maximum strength 
$\epsilon \rightarrow \epsilon_\infty$.


{}

\onecolumn
\begin{figure}
\centerline{\psfig{file=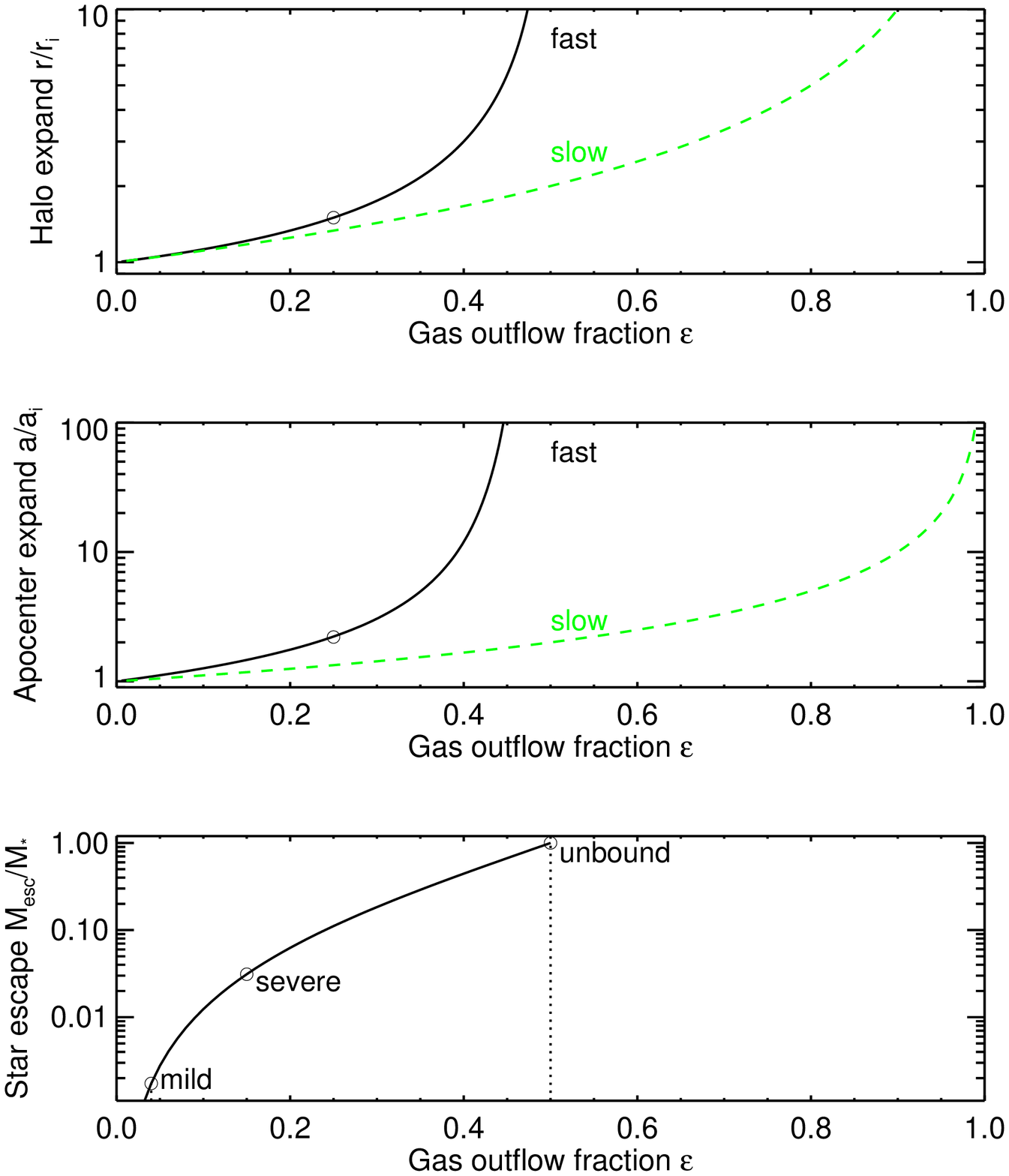}}
\label{fig1.ps}
\caption{Upper panel shows the halo expansion factor
$X=r/r_i$ in case of a slow (dashed) or a fast wind (solid) 
lossing $\epsilon$ fraction of the original mass of the galaxy everywhere.  
Middle panel shows the orbital expansion factor $A=a/a_i$ vs. $\epsilon$.
Bottom panel shows the fraction $\epsilon_*$ of all stars 
escaping from the galaxy.  
}
\end{figure}

\onecolumn
\begin{figure}
\centerline{\psfig{file=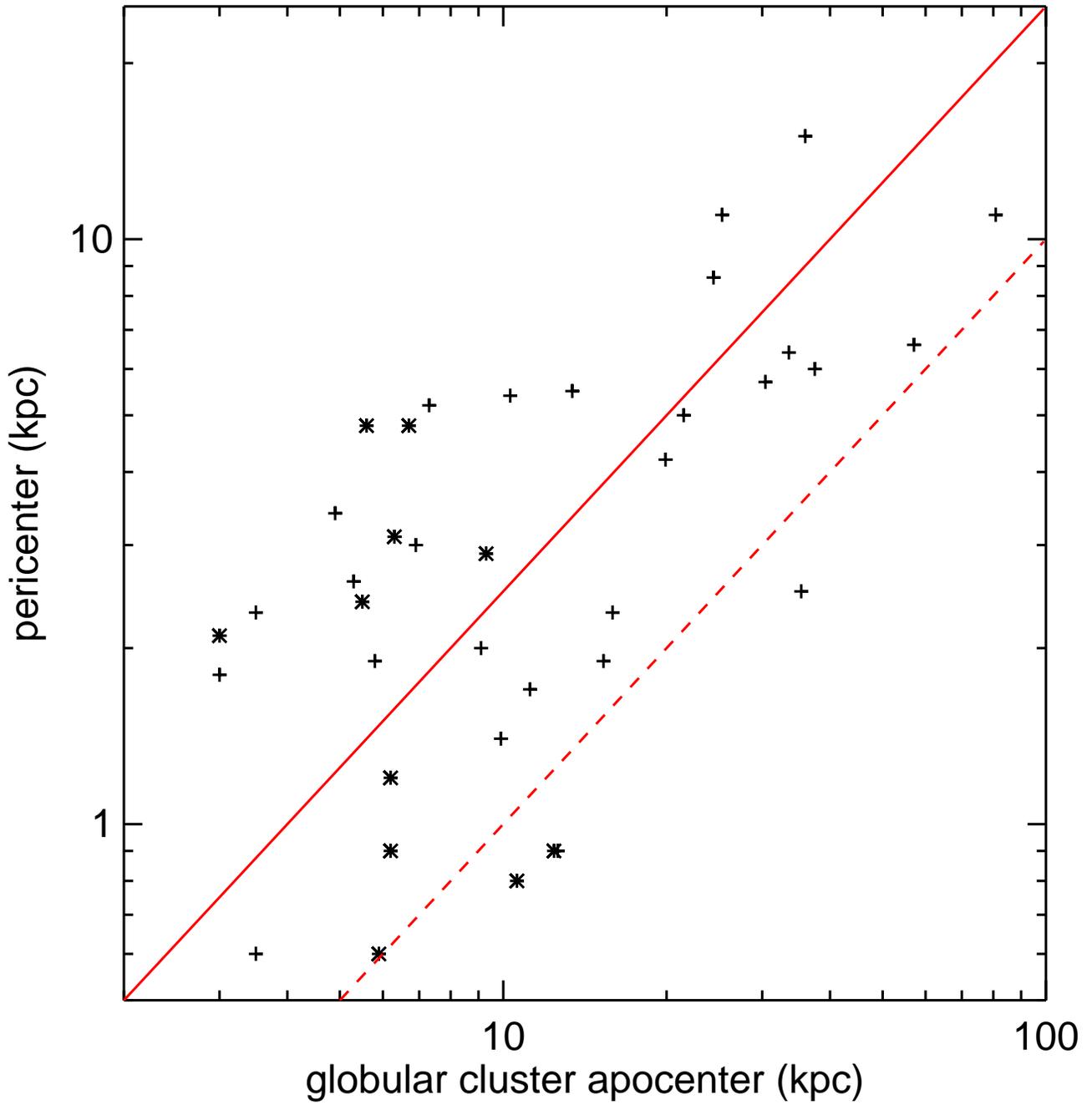}}
\label{fig2.ps}
\caption{
shows the distribution of the orbital parameters of halo globular clusters.
Disk globulars are marked with starry symbols.
The solid line indicates the predicted median apo-to-pericenter ratio 
in an isotropic model (van den Bosch et al. 1999), and the objects to the lower
right of the dashed line have an extreme apo-to-peri ratio of $\ge 10$.
}
\end{figure}

\onecolumn
\begin{figure}
\centerline{\psfig{file=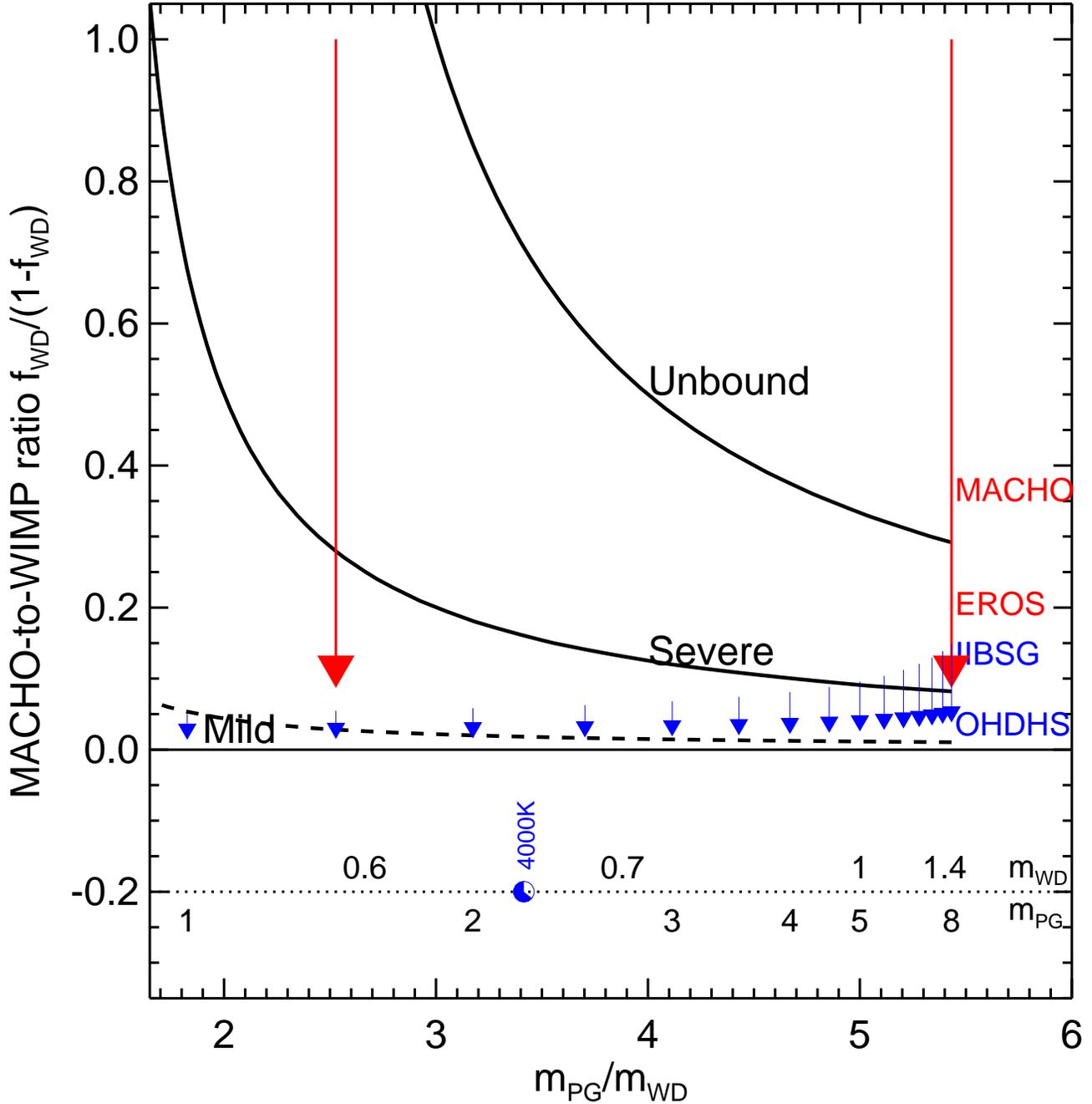}}
\label{fig3.ps}
\caption{
shows the effect of wind in the parameter
space of MACHO-to-WIMP mass ratio $f_{WD}/(1-f_{WD})$ vs. 
the progenitor-to-WD mass ratio $m_{PG}/m_{WD}$.
Models between the two solid lines can have their halo size more than doubled
by a fast wind.  Models below the dashed line are allowed.
The arrows show observational limits; the lower limits come from 
the number density by OHDHS (Oppenheimer et al. 2001), 
and IIBSG (Ibata et al. 2000); the upper limits come from the mass density by
MACHO (Alcock et al. 1997, 2000) and EROS (Lasserre et al. 2000).
A conversion scale between $m_{PG}$ and $m_{WD}$ is shown
together with a cosmic ``clock'' marker 
indicating a 4000 Kelvin WD remnant from a 10-Gyr-old burst;
hotter WDs to the left of the marker, and cooler WDs to the right.
}
\end{figure}

\onecolumn
\begin{figure}
\centerline{\psfig{file=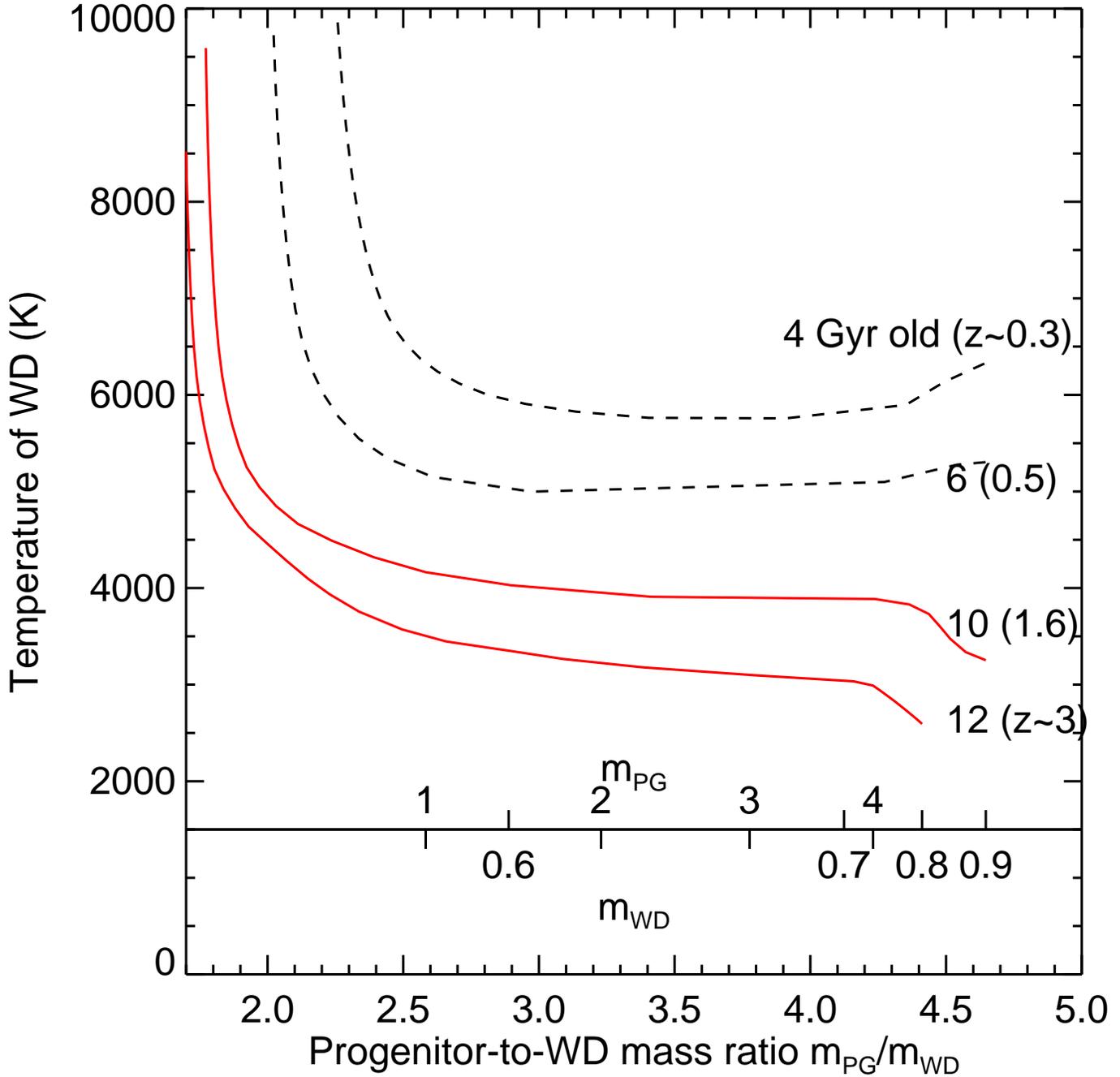}}
\label{fig4.ps}
\caption{
shows the temperature of WDs, which are predicted with 
the cooling isochrones of Richer et al. (2000) for hydrogen-rich 
WDs, together with a scale showing the initial-final mass conversion.  
Each isochronic curve is labeled
with the age of the burst and corresponding redshift in bracket.}
\end{figure}

\onecolumn
\begin{figure}
\centerline{\psfig{file=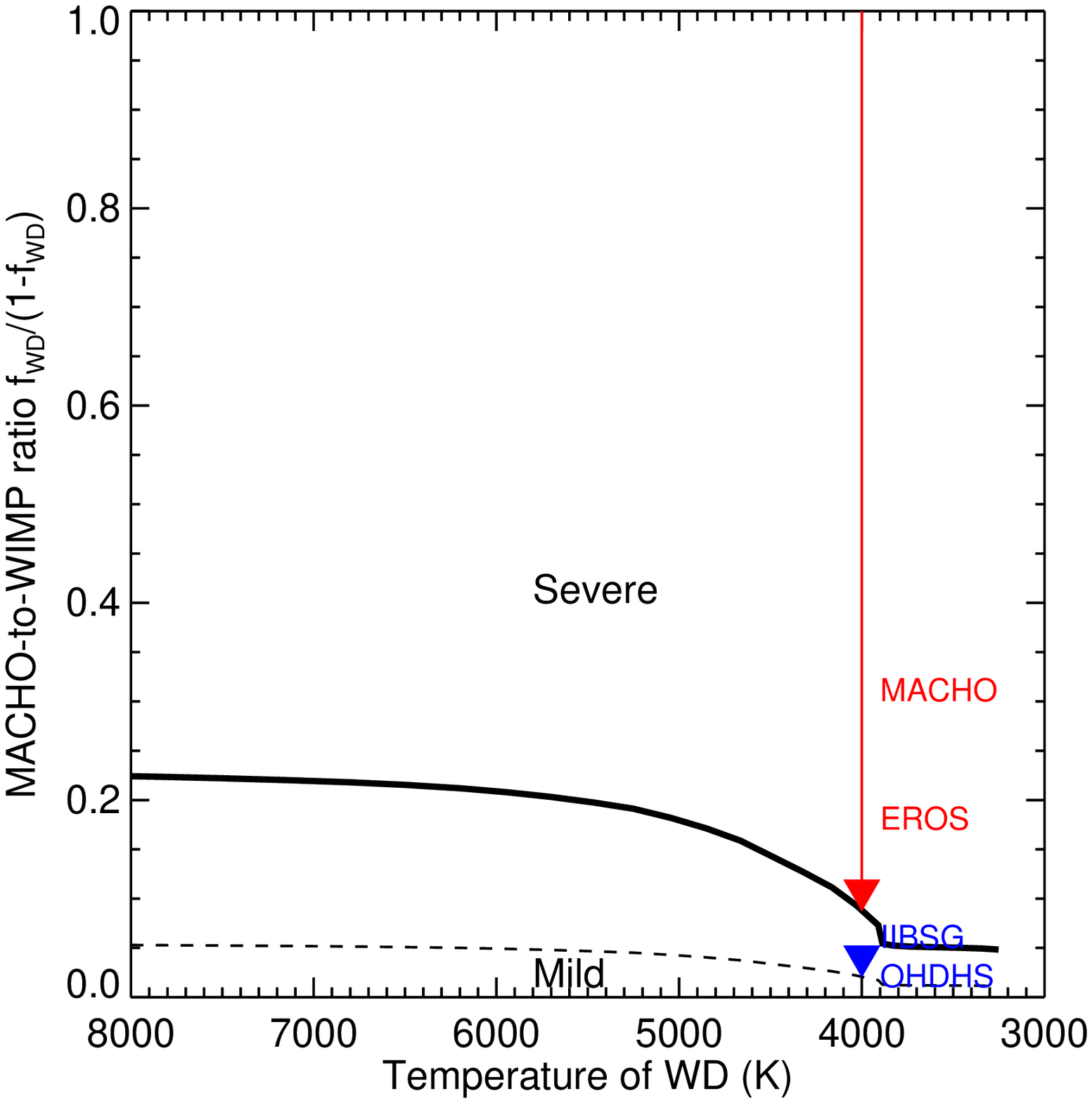}}
\label{fig5.ps}
\caption{
shows the allowed halo WD and WIMP mix ratio $f_{WD}/(1-f_{WD})$ 
and the temperature of WDs.
The region above the solid line is excluded because of too much mass
loss from the WD progenitors which could double the size of the halo.
The dashed line corresponds to a mild expansion of the halo.
Here we take a 10 Gyr cooling isochrone and the initial-final relation from
Richer et al. (2000, and references therein).  The dynamical limit
comes from eq.~(\protect{\ref{eps}}) and eq.~(\protect{\ref{mild}}).
Several estimations for $f_{WD}$ from different observations 
are also roughly indicated by the labels and arrows
at 4000 K, the typical temperature of observed WDs.}
\end{figure}

\vfill \eject
\label{lastpage}

\begin{thebibliography}{}
\bibitem{} Alcock C. et al. (MACHO collaboration) 1997, ApJ 486, 697
\bibitem{} Alcock C. et al. (MACHO collaboration) 2000, ApJ, 542, 281 
\bibitem{} Alcock C. et al. (MACHO collaboration) 2001, ApJ, 550, L169
\bibitem{} Barmby P. \& Huchra J.P. 2002, AJ in press, astro-ph/0201253 
\bibitem{} Bookbinder J., Cowie L.L., Ostriker J.P., Krolik J.H., Rees M. 1980, ApJ, 237, 648
\bibitem{} Bullock J. S. et al., 2001, MNRAS, 321, 559
\bibitem{} Carr B, 1994, ARA\&A, 32, 531 
\bibitem{} Chabrier, G., Segretain, L., and M\'era, D., 1996, ApJ, 468, L21
\bibitem{} Dinescu D., Girard T. M., \& van Altena W. F., 1999, AJ, 117, 1492
\bibitem{} Faulkner D.J. \& Freeman K.C. 1977, ApJ, 211, 77 
\bibitem{} Fields B. D., Freese K, Graff D., 2000, ApJ, 534, 265
\bibitem{} Fields, B. D., Mathews, G. J., \& Schramm, D. N. 1997, ApJ, 483, 625
\bibitem{} Gnedin O. \& Zhao H.S. 2001, MNRAS, in press
\bibitem{} Hansen B., 1999, ApJ, 520, 680
\bibitem{} Hansen B., 2001, ApJ, 558, L39
\bibitem{} Heckman T.M. 2001, ``Galactic Superwinds Circa 2001'' in ``Extragalactic Gas at Low Redshift'', ed. J. Mulchaey and J. Stocke, ASP Conf. Series (astro-ph/0107438)
\bibitem{} Hernanz, M., Garc\'ia-Berro, E., Isern, J., Mochkovitch, R., Segretain, L., \& Chabrier, G. 1994, ApJ, 434, 652
\bibitem{} Ibata R., Irwin M.J., Bienaym\'e O., Scholz R., Guibert J., 2000, ApJ, 532, L41
\bibitem{} Knox, R. A., Hawkins, M. R. S., \& Hambly, N. C. 1999, MNRAS, 306, 736 
\bibitem{} Koopmans L. \& Blanford R. 2001, astro-ph/0107358
\bibitem{} Lacey C.G. \& Ostriker J.P. 1985, ApJ, 299, 633
\bibitem{} Lasserre T. et al. (EROS collaboration) 2000, A\&A, 355, L39
\bibitem{} Lynden-Bell D. 1963, The Observatory, 83, 23
\bibitem{} Madau P. \& Rees M.J. 2001, ApJ, 551, L27
\bibitem{} Moore B. et al. 1998, ApJ, 499, L5
\bibitem{} Moore B. et al. 1999, ApJ, 304, 465
\bibitem{} Navarro J.F., Frenk C.S., \& White S.D.M., 1996, ApJ, 462, 563
\bibitem{} Oppenheimer, B.R., Hambly, N.C., Digby, A.P., Hodgkin, S.T., \& Saumon, D., Science, Vol. 292, 698
\bibitem{} Oswalt, T. D., Smith, J. A., Wood, M. A., \& Hintzen, P. 1996, Nature, 382, 692 
\bibitem{} Pettini M., Rix S.A., Steidel C.C., Adelberger K. L., Hunt M. P., \& Shapley A.E. 2002, ApJ, in press (astro-ph/0110637)
\bibitem{} Reid N., Sahu K. \& Hawley S. 2001, ApJ, 559, 942
\bibitem{} Richer H B., 2001, in proceedings of ``The Dark Universe: Matter, Energy and Gravity" (astro-ph/0107079)
\bibitem{} Richer H B., Hansen B., Limongi M, Chieffi A, Straniero O, Fahlman G. G. 2000, ApJ, 529, 318
\bibitem{} Sahu K. 1994, Nature, 370, 275
\bibitem{} Salaris, M., Degl'Innocenti, S., \& Weiss, A. 1997, ApJ, 484, 986
\bibitem{} Theuns T., Warren S. 1997, MNRAS, 284, L11
\bibitem{} van den Bosch F. C., Robertson B. E., Dalcanton J. J., de Blok W. J. G., 2000, AJ, 119, 1579
\bibitem{} van den Bosch F. C., Lewis G., Lake G. \& Stadel J. 1999, 515, 50
\bibitem{} Wilkinson M. \& Evans N.W., 1999, MNRAS, 310, 645
\bibitem{} Zhao H.S. \& Evans N.W., 2000, ApJ, 545, L35
\end{thebibliography}
\end{document}